\documentstyle[12pt]{article}
\newcommand{\be}{\begin{equation}}
\newcommand{\ee}{\end{equation}}
\newcommand{\beqas}{\begin{eqnarray*}}
\newcommand{\eeqas}{\end{eqnarray*}}
\newcommand{\beqar}{\begin{eqnarray}}
\newcommand{\eeqar}{\end{eqnarray}}

\begin{document}
\title{Worse fluctuation method for fast Value-at-Risk estimates}

\author{Jean-Philippe
Bouchaud$^{\dagger,*}$ and Marc Potters$^\dagger$}
\date{{\small $^\dagger$ Science \& Finance, 109-111 rue Victor Hugo,
92532
Levallois {\sc cedex}, FRANCE;\\ http://www.science-finance.fr\\
$^*$ Service de Physique de l'\'Etat Condens\'e,
 Centre d'\'etudes de Saclay,\\
Orme des Merisiers,
91191 Gif-sur-Yvette {\sc cedex}, FRANCE\\}
\today}
\maketitle
\begin{abstract}
We show how one can actually take advantage of the strongly
non-Gaussian nature of the fluctuations of financial assets to
simplify the calculation of the Value-at-Risk of complex non linear
portfolios.  The resulting equations are not hard to solve
numerically, and should allow fast VaR and $\Delta$VaR estimates of
large portfolios, where {\it by construction} the influence of rare
events is taken into account reliably. Our method can be seen as a
correctly probabilized `scenario' calculation (or `stress-testing').
\end{abstract}

\vskip 1cm

\subsection{Introduction}

A very important issue for the control of risk of complex portfolios, which
involves many non linear assets such as options, interest rate derivatives, etc.
is to be able to estimate reliably its Value-at-Risk, or equivalently the probability
of large downward moves, deep in the tails of the probability distributions \cite{VaR,Book}.
This is a difficult problem, since both the non-Gaussian nature
of the fluctuations of the underlying assets and the non-linear dependence of the price of the derivatives must be dealt
with. A solution to cope with non-linearity 
is to use Monte-Carlo simulations based on Gaussian multivariate
statistics for the time evolutions of all the assets underlying the portfolio. This solution is
however time consuming (specially to obtain good statistics in the tails
of the distribution) and cannot be used for fast VaR, or $\Delta$VaR calculations, 
which are important for real time estimates of the influence of a particular trade on the
global exposure of a portfolio. More importantly, this method is not reliable because of the
strongly non-Gaussian nature of the extreme moves. `Fat tails' effects are well known and lead to
a significant increase of the VaR estimate, even in the simplest case of a linear portfolio, for example containing
stocks only. These fat tails can be further amplified by the non linear nature of the relation 
between derivative products and the underlying assets, thereby leading to very large 
differences between a Gaussian VaR estimate and reality.

The aim of this paper is to introduce a method, called the `optimal fluctuation method'
in the physics literature \cite{pastur}. This method is well suited to estimate large risks
in the case where the fluctuations of the `explicative factors' are strongly
non-Gaussian (a more precise statement will be made below).  An approximate formula can
be obtained for the Value-at-Risk of a general non linear portfolio. This formula can easily be implemented
numerically. The basic idea is to identify the `most dangerous' market configuration for a given portfolio.
In the case of strongly non Gaussian fluctuations, the largest moves of the portfolio correspond to a
large change in one explicative factor, accompanied by the simultaneous `typical' changes of all the others.
In a Gaussian world, on the opposite, the large moves of a portfolio correspond to a `conspiracy', where all
factors coherently change by a small amount. 

\subsection{Fat-tailed explicative factors}

Let us assume that the variations of the value of the portfolio can be written as a 
function $df(e_1,e_2,...,e_M)$ of a set of $M$ independent random variables $e_a$, $a=1,
...,M$, called `explicative factors'. These factors can be determined by a classical Principal Component
Analysis, where the correlation matrix of all assets' increments is diagonalized. However, it might be more useful for
our purpose to consider other definitions of the correlation, more suited to tail events (see, e.g. \cite{Book}). On
short time scales, most relevant for VaR estimates, all
trend effects are negligible, and we shall therefore set $\langle e_a \rangle=0$ and $\langle e_a e_b \rangle=
\delta_{a,b} \sigma_a^2$, where $\langle ... \rangle$ denotes the average over the relevant probability distribution.
The sensitivity of the portfolio to these `explicative factors' 
can be
measured as the derivatives of the value of the portfolio with respect to the
$e_a$. We shall therefore introduce the different $\Delta$'s and $\Gamma$'s as:
\be
\Delta_a = \frac{\partial f}{\partial e_a} \qquad
\Gamma_{a,b} = \frac{\partial^2 f}{\partial e_a \partial e_b}.
\ee

We are interested in the probability for a large fluctuation $df^*$ of the
portfolio. We will first surmise that this is due to a particularly large fluctuation
of one explicative factor -- say $a=1$ -- that we will call the `dominant' factor. (The generalization to several
factors will be discussed below). Note that this is not always true, 
and depends on the statistics of the fluctuations of the $e_a$.
A condition for this assumption to be true will be discussed below, and requires
in particular that the tail of the dominant factor should not decrease faster 
than an exponential. Fortunately, this is a good assumption in financial markets, but would be completely wrong for
Gaussian statistics. Note also that the dominant
factor depends {\it a priori}, via the $\Delta$'s, on the portfolios composition.

\subsection{The dominant factor approximation}

The aim is to compute the Value-at-Risk of a certain portfolio. This is defined as
the value $df^*$ such that the probability ${\cal P}_>(df^*)$  (defined as the cumulative probability that the variation
of $f$ exceeds $df^*$)
is equal to a certain probability $p$ -- say $1 \%$ for a $99\%$ confidence VaR. Our assumption about the existence of a 
dominant factor means that these events
correspond to a market configuration where the fluctuation $\delta e_1$ is large, while all other factors are relatively
small.
Therefore, the large
variations of the portfolio can be approximated as:
\be
df(e_1,e_2,...,e_M) = df(e_1) + \sum_{a=2}^M \Delta_a e_a + \frac{1}{2} \sum_{a,b=2}^M \Gamma_{a,b}  e_a e_b,
\ee
where $df(e_1)$ is a shorthand notation for $df(e_1,0,....0)$, 
and all the derivatives are calculated at the point $(e_1,0,....0)$.
Now, we use the fact that:
\be
\label{varealpha}
{\cal P}_>(df^*)=\int \prod_{a=1}^M de_a \  P(e_1,e_2,...,e_M) 
\Theta(df(e_1,e_2,...,e_M)-df^*),
\ee
where $\Theta(x>0)=1$ and $\Theta(x<0)=0$ is the Heaviside function. We now expand the $\Theta$ function to second
order, and perform
the integration over the $e_a$'s ($a>1$), to finally obtain (see \cite{Book} for details):

\be
\label{varealpha2}
{\cal P}_>(df^*)={\cal P}_>(e_1^*) + \sum_{a=2}^M \frac{\Gamma^*_{a,a} \sigma_a^2}{2 \Delta_1^*} P(e_1^*)
- \sum_{a=2}^M \frac{\Delta_a^{*2} \sigma_a^2}{2\Delta_1^{*2}} \left(P'(e_1^*)+ \frac{\Gamma^*_{1,1}}{\Delta^*_1}
P(e_1^*)\right),
\ee
 where $e_1^*$ is such that $df(e_1^*)=df^*$, and $\Delta_1^*$ is computed
for $e_1=e_1^*$, $e_{a>1}=0$ and where $P(e_1)$ is the marginal probability distribution of the first factor. This
probability density can be estimated empirically, and fitted to one of the possible 
distribution known to describe well financial data, such as a Truncated L\'evy, Hyperbolic, or
Student distribution \cite{Book}.

In order to find the Value-at-Risk $df^*$, one should thus solve (\ref{varealpha2}) for $e_1^*$ 
with ${\cal P}_>(df^*)=p$, and then compute $df(e_1^*,0,...,0)$. Note that the equation
is not trivial since the Greeks must be estimated at the solution point $e_1^*$. In practice, the dominant factor
can be found by trial and error, by computing $e_a^*$ for all $M$ explicative factors, and choosing the
one that leads to the largest VaR.

\subsection{Discussion of the result}

Let us discuss the general result (\ref{varealpha2}) in the simple case of a
linear portfolio of assets, such that no convexity is present: the $\Delta_a$'s are constant and the $\Gamma_{a,a}$'s
are all zero.  
The equation then takes the following simpler form:
\be
{\cal P}_>(e_1^*) - \sum_{a=2}^M \frac{\Delta_a^2 \sigma_a^2}{2\Delta_1^2} P'(e_1^*) = p.
\ee
Naively, one could have thought that in the dominant factor approximation, the value
of $e_1^*$ would be the Value-at-Risk value of $e_1$ for the probability $p$, defined as:
\be
\label{evar}
{\cal P}_>(e_{1,VaR})=p.
\ee
However, the above equation shows that there is a correction term proportional to
$P'(e_1^*)$. Since the latter quantity is negative, one sees that $e_1^*$ is actually
larger than $e_{1,VaR}$, and therefore $df^* > df(e_{1,VaR})$. This
reflects the effect of all other factors, which tend to increase the Value-at-Risk
of the portfolio. 

The result obtained above relies on a second order expansion; when are higher order
corrections negligible? It is easy to see that higher order terms involve higher
order derivatives of $P(e_1)$. A condition for these terms to be negligible 
in the limit $p \to 0$, or $e_1^* \to \infty$, is that the successive derivatives 
of $P(e_1)$ become smaller and smaller. This is true provided
that $P(e_1)$ decays more slowly than exponentially, for example as a power-law. In this case, which corresponds
to financial reality \cite{tails1,tails2,Book}, each term in the expansion is a factor $1/e_1^*$ smaller than the 
previous one, which indeed
becomes negligible in the limit $p \to 0$, $e_1^* \to \infty$. On the contrary, when $P(e_1)$ 
decays faster than exponentially (for example in the Gaussian case), then the 
expansion proposed above completely looses its meaning, since higher and higher corrections become dominant when $p \to
0$. 
This is indeed expected: in a 
Gaussian world, a large event results from the accidental superposition of many 
small events; the idea of expanding around one large even is therefore not adapted.  In a power-law world, large events
are 
associated to one single
large fluctuation which dominates over all the others, and the above method is congenial for fast and precise VaR
estimates. 
The limiting case where $P(e_1)$ decays
as an exponential is interesting, since it is often a good approximation for the tail of the fluctuations of financial
assets.
Taking $P(e_1) \simeq \alpha_1 \exp -\alpha_1|e_1|$, one finds that $e_1^*$ is the solution of:
\be
e_1^* = \frac{1}{\alpha_1} \left[\log \frac{1}{p} + \log\left(1+\sum_{a=2}^M \frac{\Delta_a^2 \alpha_1^2
\sigma_a^2}{2\Delta_1^2}\right)\right].
\ee
Since one has $\sigma_1^2 \propto \alpha_1^{-2}$, the correction term is small provided
that the variance of the portfolio generated by the dominant factor is much larger
than the sum of the variance of all other factors. 

Coming back to equation (\ref{varealpha2}), one expects that if the dominant factor
is correctly identified, and if the distribution is such that the above expansion makes sense, an approximate solution
is given by $e_1^*=e_{1,VaR}+\epsilon$, with:
\be
\epsilon \simeq   \sum_{a=2}^M \frac{\Gamma_{a,a} \sigma_a^2}{2 \Delta_1}
- \sum_{a=2}^M \frac{\Delta_a^{2} \sigma_a^2}{2\Delta_1^{2}} \left(\frac{P'(e_{1,VaR})}{P(e_{1,VaR})}+
\frac{\Gamma_{1,1}}{\Delta_1}\right),
\ee
where now all the Greeks at estimated at $e_{1,VaR}$.

\subsection{Numerical tests and improved approximation schemes}

We have tested numerically the above idea by calculating the VaR of a
simple portfolios which depend on four independent factors
$e_1,...,e_4$, that we chose to be independent Student variables of
unit variance with a tail exponent $\mu=4$. We have considered two
different portfolios, `linear' ({\cal L}), such that its variations
are given by $d{\cal L}=e_1 + \frac{1}{2} e_2 + \frac{1}{5} e_3
+ \frac{1}{20} e_4$, and
`quadratic' ({\cal Q}), such that its variations are given by $d{\cal
Q}\equiv d{\cal L}+(d{\cal L})^2$. We have determined the $99 \%$, $99.5 \%$ and
$99.9 \%$ VaR of these portfolios both using a Monte-Carlo ({\sc mc})
scheme and the above approach. The results are summarized in Table 1,
where we give the simple estimate based on the VaR on the most
dangerous factor $e_{1,VaR}$, (called Th.\ 0) and the estimate which
includes the contribution of the other factors $e_1^*$ (called Th.\
1). For both {\cal L} and {\cal Q}, one sees that the latter estimate
represents a significant improvement over the naive $e_{1,VaR}$
estimate. One the other hand, our calculation still underestimates the
{\sc mc} result, in particular for the {\cal Q} portfolio. This can be
traced back to the fact that there are actually other different
dangerous market configurations which contribute to the VaR for this
particular choice of parameters. Our formalism can however easily be
adapted to the case where two (or more) dangerous configurations need
to be considered. The general equations read:
\be
\label{varealpha2bis}
{\cal P}_{>a}={\cal P}_>(e_a^*) + \sum_{b \neq a}^M \frac{\Gamma^*_{b,b} \sigma_b^2}{2 \Delta_a^*} P(e_a^*)
- \sum_{b \neq a}^M \frac{\Delta_b^{*2} \sigma_b^2}{2\Delta_a^{*2}} \left(P'(e_a^*)+ \frac{\Gamma^*_{a,a}}{\Delta^*_a}
P(e_a^*)\right),
\ee
where $a=1,...,K$ are the $K$ different dangerous factors. The $e_a^*$, and therefore $df^*$,  
are determined by the following $K$ conditions:
\be
df^*(e_1^*)=df^*(e_2^*)=...=df^*(e_K^*)\qquad {\cal P}_{>1}+{\cal P}_{>2}+...+
{\cal P}_{>K}=p.
\ee
We have computed the VaR for the different portfolios in the $K=2$
approximation. The results are reported in Table 1 under the name
`Th.\ 2'; one sees that the agreement with the {\sc mc} simulations is
improved, and is actually excellent in the linear case, where both the
configurations $e_1$ positive and large and $e_2$ positive and large
contribute. In the quadratic case, the two most dangerous
configurations are $e_1$ positive and large and $e_1$ negative and
large. In order to get perfect agreement with the Monte-Carlo result,
one should extend the calculation to $K=3$, taking into account the
configuration where $e_2$ positive and large.

\begin{table}
\begin{center}
\begin{tabular}{||l||c|c||c|c||c|c||}\hline\hline
&\multicolumn{2}{c||}{$p=1\%$}&\multicolumn{2}{c||}{$p=0.5\%$}&
\multicolumn{2}{c||}{$p=0.1\%$}
\\
& ${\cal L}$ & ${\cal Q}$ & ${\cal L}$ & ${\cal Q}$ &  ${\cal L}$ & ${\cal Q}$  \\\hline\hline
{\sc mc} &         2.93 &   13.3 &  3.53 & 18.7 & 5.30 & 40.6 \\
Th.\ 0    &         2.65&  9.7 & 3.26 & 13.9& 5.07 & 30.8\\
Th.\ 1    &         2.83&  10.9 & 3.42 & 15.1& 5.20 & 32.2\\
Th.\ 2    &      2.93&   12.1 & 3.52  & 17.2& 5.30 & 38.6 \\\hline\hline
\end{tabular}
\end{center}
\caption{}
\end{table}
\vskip 1cm

\subsection{Conclusion}

In summary, we have shown how one can actually take advantage of the
strongly non-Gaussian nature of the fluctuations of financial assets
to simplify the analytic calculation of the Value-at-Risk of
complicated non linear portfolios.  The resulting equations are not
hard to solve numerically, and should allow fast VaR and $\Delta$VaR
estimates of large portfolios, where {\it by construction} the
influence of rare events is taken into account reliably. This method
is a short-cut to Monte-Carlo methods, in the sense that we directly
identify the events which are most relevant for extreme risks, without
having to `wait' for them to appear in a Monte-Carlo
sampling. Interestingly, the calculation allows one to visualize
directly the `dangerous' market configurations which corresponds to
these extreme risks, by re-expressing all the real assets variations
in terms of the dominant factors. In this sense, our method
corresponds to a correctly probabilized `scenario' calculation.

\subsubsection*{Acknowledgments}
We thank J.P. Aguilar, P. Cizeau, L. Laloux, A. Miens and A.
Matacz for enlightening discussions. This work has been supported by the
software company {\sc atsm}, with which this idea is currently being
implemented.

\end{document}